\documentclass[aps,prd,twocolumn,nofootinbib]{revtex4-1}

\usepackage{graphicx}

\begin{document}

\title{Application of the Principle of Maximum Conformality to Top-Pair Production}

\author{Stanley J. Brodsky$^{1}$}
\email[email:]{sjbth@slac.stanford.edu}

\author{Xing-Gang Wu$^{1,2}$}
\email[email:]{wuxg@cqu.edu.cn}

\address{$^{1}$ SLAC National Accelerator Laboratory, 2575 Sand Hill Road, Menlo Park, CA 94025, USA\\
$^{2}$ Department of Physics, Chongqing University, Chongqing 401331, P.R. China}

\date{\today}

\begin{abstract}

A major contribution to the uncertainty of finite-order perturbative QCD predictions is the perceived ambiguity in setting the renormalization scale $\mu_r$. For example, by using the conventional way of setting $\mu_r \in [m_t/2,2m_t]$, one obtains the total $t \bar{t}$ production cross-section $\sigma_{t \bar{t}}$ with the uncertainty $\Delta \sigma_{t \bar{t}}/\sigma_{t \bar{t}}\sim \left({}^{+3\%}_{-4\%}\right)$ at the Tevatron and LHC even for the present NNLO level.

The Principle of Maximum Conformality (PMC) eliminates the renormalization scale ambiguity in precision tests of Abelian QED and non-Abelian QCD theories. By using the PMC, all non-conformal $\{\beta_i\}$-terms in the perturbative expansion series are summed into the running coupling constant, and the resulting scale-fixed predictions are independent of the renormalization scheme. The correct scale-displacement between the arguments of different renormalization schemes is automatically set, and the number of active flavors $n_f$ in the $\{\beta_i\}$-function is correctly determined.

The PMC is consistent with the renormalization group property that a physical result is independent of the renormalization scheme and the choice of the initial renormalization scale $\mu^{\rm init}_r$. The PMC scale $\mu^{\rm PMC}_r$ is unambiguous at finite order. Any residual dependence on $\mu^{\rm init}_r$ for a finite-order calculation will be highly suppressed since the unknown higher-order $\{\beta_i\}$-terms will be absorbed into the PMC scales' higher-order perturbative terms. We find that such renormalization group invariance can be satisfied to high accuracy for $\sigma_{t \bar{t}}$ at the NNLO level.

In this paper we apply PMC scale setting to predict the $t \bar t$ cross-section $\sigma_{t\bar{t}}$ at the Tevatron and LHC colliders. It is found that $\sigma_{t\bar{t}}$ remains almost unchanged by varying $\mu^{\rm init}_r$ within the region of $[m_t/4,4m_t]$. The convergence of the expansion series is greatly improved. For the $(q\bar{q})$-channel, which is dominant at the Tevatron, its NLO PMC scale is much smaller than the top-quark mass in the small $x$-region, and thus its NLO cross-section is increased by about a factor of two. In the case of the $(gg)$-channel, which is dominant at the LHC, its NLO PMC scale slightly increases with the subprocess collision energy $\sqrt{s}$, but it is still smaller than $m_t$ for $\sqrt{s}\lesssim 1$ TeV, and the resulting NLO cross-section is increased by $\sim 20\%$. As a result, a larger $\sigma_{t\bar{t}}$ is obtained in comparison to the conventional scale setting method, which agrees well with the present Tevatron and LHC data. More explicitly, by setting $m_t=172.9\pm 1.1$ GeV, we predict $\sigma_{\rm Tevatron,\;1.96\,TeV} = 7.626^{+0.265}_{-0.257}$ pb, $\sigma_{\rm LHC,\;7\,TeV} = 171.8^{+5.8}_{-5.6}$ pb and $\sigma_{\rm LHC,\;14\,TeV} = 941.3^{+28.4}_{-26.5}$ pb.  \\

\begin{description}

\item[PACS numbers] 12.38.Aw, 14.65.Ha, 11.15.Bt, 11.10.Gh
\item[Keywords] PMC, Renormalization Scale, Top-quark pair production

\end{description}

\end{abstract}

\maketitle

\section{Introduction}

Physical predictions in Quantum Chromodynamics (QCD) are in principle invariant under any choice of renormalization scale and renormalization scheme. It is common practice to simply guess a renormalization scale $\mu_r = Q$ and keep it fixed during the calculation, $Q$ being the typical momentum transfer of the process, and then vary it over the range $[Q/2, 2Q]$ to show the scale-uncertainty \footnote{It is often argued that by varying the renormalization scale, one can estimate contributions from higher-order terms. However, this procedure only exposes the  $\{\beta_i\}$-dependent non-conformal terms,  not the entire perturbative series. }. However, this procedure leads to scheme-dependent theoretical predictions at any finite order in perturbation theory, and it gives the wrong result when applied to QED processes. In the case of the $W$ plus three-jet production at the Tevatron and LHC, this procedure can even predict negative QCD cross sections at the next-to-leading-order~\cite{wjet} because of uncancelled large logarithms as well as the ``renormalon" terms which diverge as ($n!\beta_i^{n}\alpha_s^n$) and can give sizable contributions.

As we shall discuss, the optimal procedure for obtaining precise QCD predictions is to choose the renormalization scale so that the result is scheme-independent at any fixed order in $\alpha_s$. The Brodsky-Lepage-Mackenzie method (BLM)~\cite{blm} and the Principle of Maximum Conformality (PMC)~\cite{pmc1,pmc2,pmc3} provide the solution to this problem. The PMC is the principle underlying BLM scale setting, and they are equivalent to each other through the PMC - BLM correspondence principle~\cite{pmc2}, so if not specifically stated, we shall treat them on equal footing.

The main idea of PMC scale setting is that, after proper procedures, all non-conformal $\{\beta_i\}$-terms in the perturbative expansion are summed into the running coupling so that the remaining terms in the perturbative series are identical to that of a conformal theory; i.e., the corresponding theory with $\{\beta_i\} \equiv \{0\}$. After PMC scale setting, the divergent ``renormalon" series with $n!$-growth does not appear in the conformal series. QCD predictions using PMC are then independent of the choice of renormalization scheme. Since renormalon terms are absent, one obtains a more convergent perturbative expansion series, and thus the full next-to-leading-order (NLO), or even the leading-order (LO) calculation, is often enough to achieve the required accuracy.

The PMC method satisfies all self-consistent conditions, including the existence and uniqueness of the scale, reflexivity, symmetry and transitivity~\cite{selfconsistence}. In particular, the transitivity property shows that the relation between any two physical observables is independent of the choice of intermediate renormalization scheme. The transitivity is thus equivalent to the renormalization group property that shows the predictions in pQCD are independent of the choice of an intermediate renormalization scheme \cite{ren1,ren3,gml}. In the limit $N_C \to 0$ at fixed $\alpha=C_F \alpha_s$ with $C_F=(N_c^2-1)/2N_c$ ~\cite{qedlimit,qed2}, the PMC method also agrees with the standard Gell Mann-Low procedure for setting the renormalization scale in Abelian QED. It should be recalled that there is no ambiguity in setting the renormalization scale in QED. In the standard Gell-Mann-Low scheme for QED, the renormalization scale is the virtuality of the virtual photon~\cite{gml}. For example, the renormalization scale for the electron-muon elastic scattering through one-photon exchange can be set as the virtuality of the exchanged photon, i.e. $\mu_r^2 = q^2 = t$. Thus, we have
\begin{equation}
\alpha(t) = \frac{\alpha(t_0)}{1 - \Pi(t,t_0)} \;\;, \label{qed}
\end{equation}
where $\Pi(t,t_0) = (\Pi(t,0) -\Pi(t_0,0))/ (1-\Pi(t_0,0)) $ which sums all vacuum polarization contributions to the dressed photon propagator. Because final result when summed up to all orders is independent of $t_0$, one can choose any initial renormalization scale $t_0$. Of course in QCD, the question is much more complicated due to its non-Abelian nature. However its scales can also be unambiguously set at each finite order by the PMC procedure.

Formally, as a starting point, one needs to introduce an initial renormalization scale $\mu^{\rm init}_r$ for PMC scale setting, which in practice can be set as a typical physical scale ($Q$) of the process, i.e. $\mu^{\rm init}_r =Q$. However, the final result will be independent of $\mu^{\rm init}_r$ after we have summed the relevant $\{\beta_i\}$-terms; i.e. for any physical observable ${\cal O}$, we have
\begin{equation}
\frac{\partial {\cal O}\left(\mu^{\rm PMC}_r\right)}{\partial \mu^{\rm init}_{r}} \equiv  0 ,
\end{equation}
where $\mu^{\rm PMC}_r$ stands for the PMC scale \footnote{This is different from the Principle of Minimum Sensitivity (PMS) \cite{pms} which chooses $\mu_r$ at the stationary point of ${\cal O}$, i.e. ${d {\cal O}(\mu^{\rm PMS}_{r})}/{d \mu^{\rm PMS}_{r}} = 0$. It is noted that PMS does not satisfy the transitivity property of the renormalization group \cite{selfconsistence}; i.e. it predicts different correlations between two physical observables when they are related through different intermediate schemes. Moreover, it gives unphysical results for 2-jet and 3-jet production in $e^+e^-$ annihilation within the small jet-invariant-mass region \cite{Kramer} since it sums the contributions not associated with the renormalization into the running coupling. }.

The essential steps for implementing the PMC are as follows: Consider the perturbative expansion for a physical quantity at an arbitrary initial scale $\mu^{\rm init}_r$, \begin{equation}\label{phyvalue}
\rho_n = {\cal C}_0 \alpha_s^p(\mu^{\rm init}_r) + \sum_{i=1}^{n}{\cal C}_i(\mu^{\rm init}_r) \alpha_s^{p+i}(\mu^{\rm init}_r)
\end{equation}
expanded to $n$-th order in $\alpha_s$. Its NLO coefficient ${\cal C}_1$ can be generally written as
\begin{eqnarray}
{\cal C}_1(\mu^{\rm init}_r) &=& {\cal C}_{10}(\mu^{\rm init}_r) + {\cal C}_{11}(\mu^{\rm init}_r) \; n_f  \nonumber\\
&=& \tilde{{\cal C}}_{10}(\mu^{\rm init}_r) + \tilde{{\cal C}}_{11}(\mu^{\rm init}_r) \; \beta_0 \;, \nonumber
\end{eqnarray}
where $\beta_0=11-2/3 n_f$, the coefficients ${\cal C}_{10}$ and ${\cal C}_{11}$ are $n_f$-independent, $\tilde{{\cal C}}_{10}={\cal C}_{10}+\frac{33}{2}{\cal C}_{11}$, and $\tilde{{\cal C}}_{11}=-\frac{3}{2}{\cal C}_{11}$. The LO PMC scale $\mu^{\rm PMC}_r$ is then set by the condition
\begin{equation}\label{pmcbasic}
\tilde{{\cal C}}_{11}(\mu^{\rm PMC}_r) = 0.
\end{equation}
This prescription ensures that, as in QED, all vacuum polarization contributions due to the light-fermion pairs are absorbed into the coupling constant. The resulting series is conformal since the nonzero $\{\beta_i\}$-terms are now absent. Moreover the PMC prediction is independent of the choice of renormalization scheme, as required by the renormalization group invariance. The specific value of the PMC scale will depend on the scheme, but the physical result is the same; i.e., the displacement between the arguments of different schemes is accounted for. Note that because ${{\cal C}}_{11} \propto \tilde{{\cal C}}_{11}$, one can in practice obtain the PMC scale by using the equation ${{\cal C}}_{11}(\mu^{\rm PMC}_r) \equiv 0$, which is usually adopted in the BLM-literature \footnote{This should be used with care, since if ${\cal C}_{10}$ is a constant free of scale, then one will  obtain the wrong NLO coefficient instead of the correct one ${\tilde{\cal C}}_{10}$. }.

The PMC scale setting fixes the final renormalization scale at any finite order and it is consistent with this renormalization group equation theorem. This is also the invariance principle used to derive the renormalization group results such as the Callan-Symanzik equations~\cite{callan,symanzik}. The physical observable ${\cal O}\left(\mu^{\rm PMC}_r\right)$ is independent of the choice of different renormalization scheme such as the $\overline{MS}$-scheme, the momentum-subtraction scheme and etc. Such scheme-independence can be adopted to derive commensurate scale relations among different observables and to find proper scale-displacements among the PMC scales which are derived under different schemes or conventions \cite{stanlu}. After PMC scale setting, the number of active flavors $n_f$ in the $\{\beta_i\}$-function is correctly determined \cite{sjbnf}.

As a fixed-order QCD calculation, there is some residual initial-scale dependence because of the unknown-higher-order $\{\beta_i\}$-terms. Such residual renormalization scale-uncertainty can be greatly suppressed due to the fact that those higher-order $\{\beta_i\}$-terms will be absorbed into the PMC scales' higher-order $\alpha_s$-terms. Recently, we have demonstrated a systematic and scheme-independent treatment of PMC up to next-to-next-to-leading order (NNLO)~\cite{pmc2}. A brief report on PMC is presented in Ref.\cite{pmc3} \footnote{Another way of treating the $\{\beta_i\}$-series to set the PMC/BLM scale can be found in Refs.\cite{Kataev,Mikhailov1}.}. As an important application, we shall in this paper predict the total top-quark pair production cross-section up to NNLO without renormalization scheme or scale ambiguity.

The top quark is the heaviest elementary particle within the Standard Model, and because of its large coupling to the Higgs, the top-quark production processes provide a sensitive probe of electroweak symmetry breaking. The precise measurement of the top-quark properties is an important task for the Tevatron and the LHC hadron colliders. The top-quark pair production cross-section can be used as a basic quantity to measure the top-quark pole mass $m_t$, to constrain new physics and to extract useful information on the parton distribution function (PDF) of the proton. Thus the top-quark production cross-section is of central and fundamental interest, both theoretically and experimentally.

The total top-pair production cross-section $\sigma_{t \bar{t}}$ has been measured at the Tevatron with a precision $\Delta \sigma_{t \bar{t}}/\sigma_{t \bar{t}}\sim\pm 7\%$~\cite{cdf,d0} and the two LHC experiments have already reached similar sensitivity~\cite{atlas,cms}. With more statistics forthcoming, a more precise theoretical prediction is required.

The total cross-section for the top-pair production has been calculated up to NLO within the $\overline{MS}$-scheme in Refs.~\cite{nason1,nason2,beenakker1,beenakker2}; explicit analytical results are provided in Ref.~\cite{czakon1}. Large logarithmic corrections associated with the soft gluon emission have been investigated and resummed up to next-to-next-to-leading-logarithmic corrections~\cite{moch1,moch3,czakon2,beneke1,nason3,andrea,vogt}. Even though the complete NNLO fixed-order results are not available at present, parts of the full NNLO fixed order results have been derived through resummation~\cite{moch1,moch2,beneke2,nik}. This NNLO approximation is supported by the observation that the production of a top-quark pair with an additional jet is small~\cite{jet1,jet2}. These results provide the foundation for approximating the NNLO results.

The dependence on the renormalization and factorization scales at the NNLO accuracy has been discussed using conventional procedures in Refs.~\cite{moch2,moch3}. In this paper, we will provide a new perspective using PMC scale setting. We shall first set the PMC scales up to NLO with the help of the NNLO analytic expressions, and we will then give a detailed discussion on how this procedure improves the predictability of pQCD. For this purpose, we shall adopt the analytical expressions up to NNLO provided in the literature, e.g. Ref.~\cite{hathor}.

The remaining parts of this paper are organized as follows: in Sec.II, we give the PMC scale setting for the top-quark pair production. In Sec.III, we present numerical results and discussions. Sec.IV provides a summary.

\section{PMC Scale setting for the top-quark pair production}

In this section, we shall give the general formulae for setting the PMC scales for the top-quark pair production cross-section up to NLO. We will then compute the scales for each production channel and discuss their specific properties.

\subsection{General formulae}

\begin{center}
\begin{figure}
\includegraphics[width=0.35\textwidth]{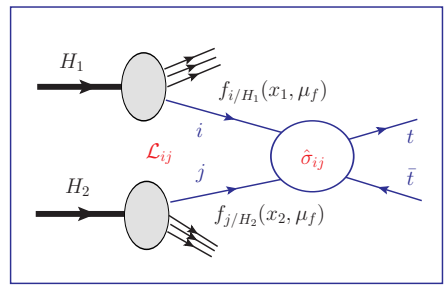}
\caption{Diagram for the top-quark pair production obtained from the convolution of the partonic subprocess cross-section $\hat \sigma_{ij}$ with the parton luminosities ${\cal L}_{ij}$. }
\label{figure}
\end{figure}
\end{center}

Hadronic cross section for the top-quark pair production, as illustrated in Fig.(\ref{figure}), can be obtained from the convolution of the factorized partonic cross-section $\hat \sigma_{ij}$ with the parton luminosities ${\cal L}_{ij}$:
\begin{equation}
\sigma_{H_1 H_2 \to {t\bar{t} X}} = \sum_{i,j} \int\limits_{4m^2_{t}}^{S}\, ds \,\, {\cal L}_{ij}(s, S, \mu_f) \hat \sigma_{ij}(s,\alpha_s(\mu_r),\mu_r,\mu_f) , \label{basic}
\end{equation}
with the parton luminosity
\begin{displaymath}
{\cal L}_{ij} = {1\over S} \int\limits_s^S {d\hat{s}\over \hat{s}} f_{i/H_1}\left(x_1,\mu_f\right) f_{j/H_2}\left(x_2,\mu_f\right),
\end{displaymath}
where $x_1= {\hat{s} / S}$ and $x_2= {s / \hat{s}}$. Here $S$ denotes the hadronic center-of-mass energy squared and $s=x_1 x_2 S$ is the subprocess center-of-mass energy squared. we denote the renormalization scale as $\mu_r$ and the factorization scale as $\mu_f$. The functions $f_{i/H_{1,2}}(x_\alpha,\mu_f)$ ($\alpha=1,2$) are the PDFs describing the probability to find a parton of type $i$ with a momentum fraction between $x_\alpha$ and $x_{\alpha} +dx_{\alpha}$ in the hadron $H_{1,2}$. The top-quark mass $m_{t}$ is the mass renormalized in the on-shell (pole-mass) scheme. Setting $s=4m_t^2 (S/4m_t^2)^{y_1}$ and $\hat{s}=s(S/s)^{y_2}$, we can transform the two-dimensional integration over $s$ and $\hat{s}$ into the integration over two variables $y_{1,2}\in[0,1]$. These integrals can be performed numerically using the VEGAS program~\cite{vegas} \footnote{An improvement of the VEGAS program to derive precise numerical results can be found in the generators BCVEGPY~\cite{bcvegpy} and GENXICC~\cite{genxicc}.}.

The partonic subprocess cross-sections $\hat \sigma_{ij}$ can be decomposed in terms of the dimensionless scaling functions $f^m_{ij}$, where $(ij) = \{(q{\bar q}), (gg), (gq), (g\bar{q})\}$ stands for the four production channels and $m=0,1,2$ stands for the LO, NLO and NNLO functions respectively \footnote{There are also $(qq)$- and $(\bar{q}\bar{q})$- channels at the NLO and NNLO orders. However, because of the large suppression coming from the parton luminosities and the phase-space (due to more jets in the final state), their cross-sections are much smaller than the $(gq)$- or $(g\bar{q})$-channel. We will not discuss these channels here. }. Up to NNLO, they takes the following form
\begin{eqnarray}
\hat\sigma_{ij} &=& {1 \over m^2_{t}} \Big\{f_{ij}^{0}(\rho,Q) a^2_s(Q) + f_{ij}^{1}(\rho,Q) a^3_s(Q) \nonumber\\
&& + f_{ij}^2(\rho,Q) a^4_s(Q) \Big\} , \label{subcs}
\end{eqnarray}
where $\rho=4m_t^2/s$, and we have, for convenience, temporarily set $\mu_f=\mu_r=Q$, $a_s(Q)=\alpha_s(Q)/\pi$. The analytic expressions of the scale functions $f_{ij}^{0,1,2}(\rho,Q)$, which shows the full renormalization and factorization scale dependence, can be directly read from the HATHOR program \footnote{Note that in the appendix of the HATHOR manual~\cite{hathor}, the scaling function $f^{(0)}_{q\bar{q},gg}$ in the functions $f^{21}_{q\bar{q},gg}$ and $f^{22}_{q\bar{q},gg}$ (which are not proportional to $n_f^2$) should be replaced by its limiting form at $\rho\to 1$, since all the coefficients of the scaling functions are fitted to per mil accuracy through such parameterization~\cite{moch3}.}.

We will apply the PMC scale setting procedure to each renormalizable hard subprocess which enters the pQCD leading-twist factorization procedure; the initial and final quark and gluon lines are taken to be on-shell so that the calculation of each hard subprocess amplitude is gauge invariant. Thus the application of PMC to hard subprocesses does not involve the factorization scale.

According to the PMC scale setting, we need to identify the explicit terms that are $n_f$- or $n^2_f$- dependent. One must be careful that only those $n_f$-terms associated with the $\{\beta_i\}$-terms will be absorbed into the running of $\alpha_s$. Coulomb-type corrections will lead to sizable contributions in the threshold region~\cite{coul1,coul2} which are enhanced by factors of $\pi$ and the PMC scale can be relatively soft for $v\to 0$ ($v=\sqrt{1-\rho}$, the heavy quark velocity). Thus the terms which are proportional to $(\pi/v)$ or $(\pi/v)^2$ will be treated separately~\cite{brodsky1}.

The NLO and NNLO scaling functions can be written as
\begin{eqnarray}
f_{ij}^{1}(\rho,Q) &=& \left[A_{1ij} + B_{1ij} n_f \right] + D_{1ij} \left(\frac{\pi}{v}\right)\\
f_{ij}^{2}(\rho,Q) &=& \left[A_{2ij} + B_{2ij} n_f + C_{2ij} n^2_f\right] +\nonumber\\
&& \left[D_{2ij} +E_{2ij} n_f \right] \left(\frac{\pi}{v}\right) + F_{2ij}\left(\frac{\pi}{v}\right)^2 .
\end{eqnarray}
Substituting these scaling functions into Eq.(\ref{subcs}), the partonic cross-section $\hat \sigma_{ij}$ becomes
\begin{eqnarray}
m_t^2 \hat\sigma_{ij} &=& A_{0ij} a^2_s(Q) + \nonumber\\
&&\left\{\left[A_{1ij} + B_{1ij} n_f \right] + D_{1ij} \left(\frac{\pi}{v}\right)\right\} a^3_s(Q) + \nonumber\\
&& \left\{\left[A_{2ij} + B_{2ij} n_f + C_{2ij} n^2_f\right] + \right.\nonumber\\
&& \left. \left[D_{2ij} +E_{2ij} n_f \right] \left(\frac{\pi}{v}\right) + F_{2ij}\left(\frac{\pi}{v}\right)^2\right\} a^4_s(Q) ,
\end{eqnarray}
where $A_{0ij} = f_{ij}^{0}(\rho,Q)$.

\begin{figure}
\includegraphics[width=0.4\textwidth]{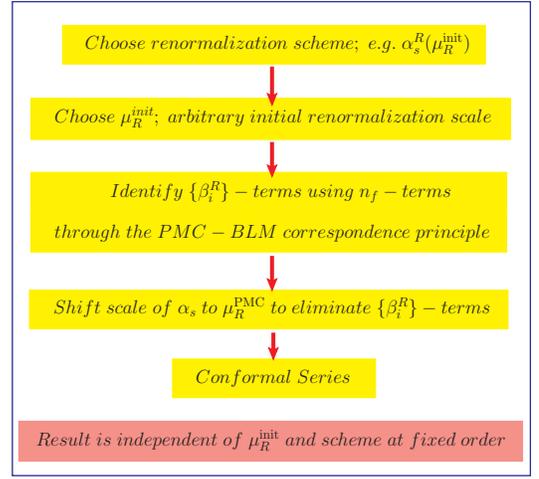}
\caption{A ``flow chart" which illustrates the PMC procedure. }
\label{fig1}
\end{figure}

Following the same procedures as described in Ref.~\cite{pmc2}, the PMC scales can be set order-by-order, treating the Coulomb and non-Coulomb contributions separately. A ``flow chart" which illustrates the PMC procedure is shown in Fig.(\ref{fig1}). More explicitly, the first step is to set the effective scale $Q^*_1$ and $Q^*_2$ at LO:
\begin{eqnarray}
m_t^2 \hat\sigma_{ij} &=& A_{0ij} a^2_s(Q_1^*) + \left[\tilde{A}_{1ij}\right] a^3_s(Q_1^*) + \nonumber\\
&& \left[\tilde{A}_{2ij} + \tilde{B}_{2ij} n_f \right]a^4_s(Q_1^*) + D_{1ij} \left(\frac{\pi}{v}\right) a^3_s(Q_2^*) +\nonumber\\
&& \left[\tilde{D}_{2ij}\right] \left(\frac{\pi}{v}\right)a^4_s(Q_2^*) + F_{2ij}\left(\frac{\pi}{v}\right)^2 a^4_s(Q_2^*) .
\end{eqnarray}
The second step is to set the effective scale $Q^{**}_1$ at NLO:
\begin{eqnarray}
m_t^2 \hat\sigma_{ij} &=& A_{0ij} a^2_s(Q_1^*) + \left[\tilde{A}_{1ij}\right] a^3_s(Q_1^{**}) + \nonumber\\
&& \left[\tilde{\tilde{A}}_{2ij} \right] a^4_s(Q_1^{**}) + D_{1ij} \left(\frac{\pi}{v}\right) a^3_s(Q_2^*) + \nonumber\\
&& \left[\tilde{D}_{2ij} \left(\frac{\pi}{v}\right) + F_{2ij}\left(\frac{\pi}{v}\right)^2 \right] a^4_s(Q_2^*) \nonumber \\
&=& A_{0ij} a^2_s(Q_1^*) + \left[\tilde{A}_{1ij}\right] a^3_s(Q_1^{**}) + \left[\tilde{\tilde{A}}_{2ij} \right] a^4_s(Q_1^{**})\nonumber\\
&& + \left(\frac{\pi}{v}\right) D_{1ij} \left[\frac{2\kappa}{1-\exp(-2\kappa)}\right] a^3_s(Q_2^*) , \label{pmceq}
\end{eqnarray}
where $\kappa=\frac{\tilde{D}_{2ij}}{D_{1ij}} a_s(Q_2^*) + \frac{F_{2ij}}{D_{1ij}} \left(\frac{\pi}{v} \right) a_s(Q_2^*)$. $Q^{*}_{1}$ and $Q^{**}_{1}$ are LO and NLO PMC scales for the usual non-Coulomb part and $Q^{*}_2$ is the LO PMC scale for the Coulomb part. Note that the effective scales will be a perturbative series of $a_s$ so as to absorb all $n_f$-dependent terms properly~\cite{pmc2,stanlu}. When performing these scale shifts, we eliminate the $n_f$-terms completely. At the same time, the corresponding coefficients at the given $\alpha_s$-order are modified, since the changes to the coefficients are proportional to $\{\beta_i\}$-functions.

For the usual non-Coulomb terms, the two PMC scales $Q^{*,**}_{1}$ satisfy
\begin{eqnarray}
\ln\frac{Q^{*2}_1}{Q^2} &=& \ln\frac{Q^{*2}_{10}}{Q^2} +\frac{\chi}{4}\beta_0 \ln\frac{Q^{*2}_{10}}{Q^2} a_s(Q) \nonumber\\
\ln\frac{Q^{*2}_{10}}{Q^2} &=& \frac{3B_{1ij}}{A_{0ij}} \;,\;\; \chi = \frac{9B^2_{1ij}-12A_{0ij}C_{2ij}}{2A_{0ij} B_{1ij}}
\end{eqnarray}
and
\begin{equation} \label{nloscale}
\ln\frac{Q^{**2}_{1}}{Q_{1}^{*2}} = \frac{2\tilde{B}_{2ij}}{\tilde{A}_{1ij}} .
\end{equation}
Here $\beta_0=\frac{11}{3}C_A -\frac{4}{3}T n_f=11-2n_f/3$, where $C_A=N_c$ and $T=1/2$ for a general $SU(N_c)$ color-group. The PMC procedure is independent of $N_c$, and here we have set $N_c=3$. The order-by-order coefficients are
\begin{eqnarray}
\tilde{A}_{1ij} &=& \frac{2A_{1ij}+33B_{1ij}}{2} \\
\tilde{A}_{2ij} &=& \frac{1}{8A_{0ij}}\left[ 99B_{1ij}(2A_{1ij}+33B_{1ij})+ \right.\nonumber\\
&& \left. A_{0ij}(8A_{2ij}+306B_{1ij}-2178C_{2ij})\right] \\
\tilde{B}_{2ij} &=& \frac{1}{4A_{0ij}}\left[4 A_{0ij}(B_{2ij}+33C_{2ij})- \right.\nonumber\\
&& \left. B_{1ij}(19A_{0ij}+6A_{1ij}+99B_{1ij})\right] \\
\tilde{\tilde{A}}_{2ij} &=& \frac{2\tilde{A}_{2ij}+33\tilde{B}_{2ij}}{2}
\end{eqnarray}

In the case of the Coulomb terms, we adopt the Sommerfeld rescattering formula to sum up the higher-order $(\pi/v)$-terms. Notice that the overall factor $({\pi}/{v})$ in the last line of Eq.(\ref{pmceq}) will be canceled by a $v^{1}$-factor in $D_{1ij}$ ensuring a finite result at $v = 0$. The LO PMC scale $Q^*_2$ satisfies
\begin{equation}
\ln\frac{Q^{*2}_{2}}{Q^2} = \frac{2E_{2ij}}{D_{1ij}}
\end{equation}
with the coefficient
\begin{equation}
\tilde{D}_{2ij} = (2D_{2ij}+33E_{2ij})/{2} .
\end{equation}

Since the channels $(ij) = \{(q{\bar q}), (gg), (gq), (g\bar{q})\}$ are distinct and non-interfering, their PMC scales should be set separately. The procedure is gauge invariant.

\subsection{$(q\bar{q})$-channel}

\begin{figure}
\includegraphics[width=0.45\textwidth]{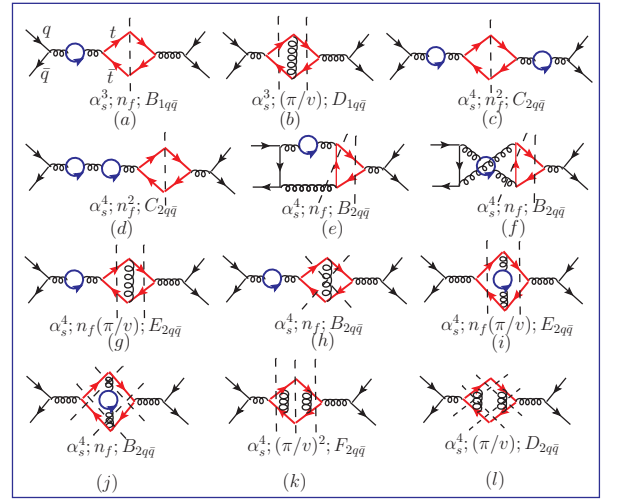}
\caption{Cut diagrams for the $n^{(1,2)}_f$-terms and the Coulomb-terms for the $(q\bar{q})$-channel up to NNLO, where the solid circles stand for the light-quark loops. }
\label{qqbar}
\end{figure}

For the $(ij)=(q\bar{q})$ channel, all the coefficients $A_{0q\bar{q}}$, $A_{1q\bar{q}}$ and etc. are non-zero. The corresponding contributions to the cross-section are conveniently expressed by the absorptive contributions (cuts) of the Feynman diagrams. We present the typical cut diagrams for the $n_f$-terms and the Coulomb-terms in Fig.(\ref{qqbar}).

\begin{figure}
\includegraphics[width=0.40\textwidth]{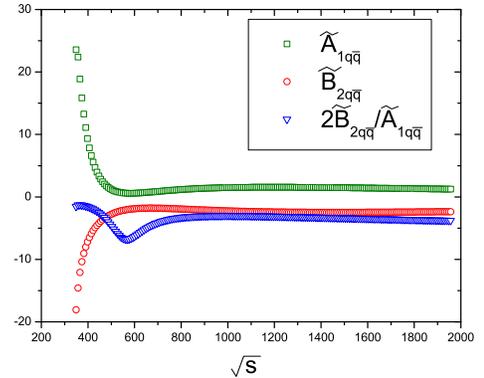}
\caption{PMC coefficients of the $(q\bar{q})$-channel versus the subprocess collision energy $\sqrt{s}$, which determine the behavior of the NLO PMC scale $Q^{**}_1$. $Q=m_t=172.9$ GeV. }
\label{qqcoe}
\end{figure}

Following PMC scale setting, all $n_f$-terms which are associated with the $\{\beta_{i}\}$-terms are absorbed into the $\alpha_s$-coupling step-by-step:
\begin{itemize}
\item When performing the scale shifts, the terms $A_{1ij}$, $A_{2ij}$, $D_{2ij}$, etc. are modified so that the final expression is conformal.
\item When performing the scale shift $Q\to Q^{*}_1$, the first type of $\{\beta_{i}\}$-terms $B_{1q\bar{q}}$ (Fig.(\ref{qqbar}a)) and $C_{2q\bar{q}}$ (Fig.(\ref{qqbar}c,\ref{qqbar}d)) are eliminated exactly. The part of $B_{2q\bar{q}}$, i.e. Fig.(\ref{qqbar}h) which contains the same type of $\{\beta_{i}\}$-term is also absorbed into $\alpha_s$-running. The remaining part of $B_{2q\bar{q}}$ (Fig.(\ref{qqbar}e,\ref{qqbar}f,\ref{qqbar}j)) is compensated by $A_{1q\bar{q}}$ and $B_{1q\bar{q}}$ to ensure that the first type of $\{\beta_{i}\}$-terms are absorbed into $\alpha_s$-coupling exactly, which results in a new variable $\tilde{B}_{2q\bar{q}}$. Because $(B_{1q\bar{q}}/A_{0q\bar{q}})$ increases monotonically with $\sqrt{s}$, the scale $Q^{*}_1$ shows the same trend versus $\sqrt{s}$.
\item When performing the scale shift $Q^{*}_1\to Q^{**}_1$, the second type of $\{\beta_{i}\}$-terms, i.e. $\tilde{B}_{2q\bar{q}}$ (Fig.(\ref{qqbar}e,\ref{qqbar}f,\ref{qqbar}j)) are eliminated. As shown in Fig.(\ref{qqcoe}), the value of $\tilde{B}_{2q\bar{q}}$ is always negative and $\tilde{A}_{1q\bar{q}}$ has a minimum value at small $\sqrt{s}$. With the help of Eq.(\ref{nloscale}), one finds that the NLO scale $Q^{**}_1$ is smaller than $Q$. Note especially that there is a concave dependence for $Q^{**}_1$ versus $\sqrt{s}$ as shown in Fig.(\ref{lhcscale}).
\item Similarly, when performing the scale shift $Q\to Q^{*}_2$ for the Coulomb type contribution, the term $E_{2q\bar{q}}$ (Fig.(\ref{qqbar}g,\ref{qqbar}i)) is eliminated.
\end{itemize}

\subsection{$(gg)$-channel}

\begin{figure}[t]
\includegraphics[width=0.40\textwidth]{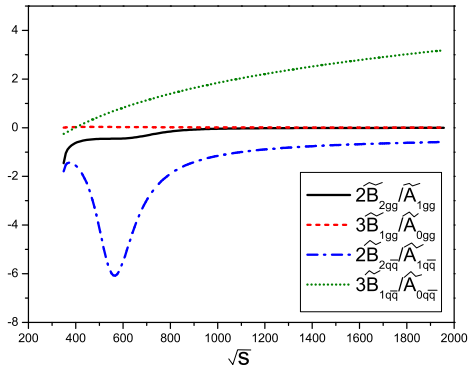}
\caption{Comparison with of the PMC coefficients for the $(gg)$- and $(q\bar{q})$- channels versus the subprocess collision energy $\sqrt{s}$. $Q=m_t=172.9$ GeV. }
\label{gqcoe}
\end{figure}

For the $(ij)=(gg)$ channel, we have $C_{2gg}=0$, and the other coefficients $A_{0gg}$, $A_{1gg}$, etc. are non-zero. Thus the $(gg)$-channel can be treated in a similar way as the $(q\bar{q})$-channel. As can be seen from Fig.(\ref{gqcoe}), in contrast to the $(q\bar{q})$-channel, both $2\tilde{B}_{2gg}/\tilde{A}_{1g\bar{g}}$ and $3B_{1gg}/A_{0g\bar{g}}$ are close to zero, thus its LO and NLO PMC scales ($Q^{*}_1$ and $Q^{**}_1$) are close to $Q$ and are changed more smoothly with $\sqrt{s}$.

In Refs.~\cite{binger,lu}, the authors have suggested another way to set the effective coupling for the three-gluon vertex which depends on its input three momenta and gives rise to an effective scale $Q^2_{eff}$ which governs the behavior of the three-gluon vertex up to one-loop level. Such effective scale is consistent with the PMC scale setting in principle, however its momenta must be space-like or time-like, and it cannot be adopted for the present case when the two initial gluon momenta are set to be on mass shell.

\subsection{$(gq)$- or $(g\bar{q})$- channel}

For the $(i,j)=(g,q)$ or $(g,\bar{q})$ channel, we have $A_{0ij}=0$, $D_{1gq}= 0$, $E_{2gq}=F_{2gq} = 0$, which shows that the Coulomb type corrections start only at the NNLO order. Thus its scale setting is greatly simplified, and we only need one LO PMC scale for these two channels, i.e.
\begin{eqnarray}
m_t^2 \hat\sigma_{ij} &=& \left[A_{1ij}\right] a^3_s(Q) + \left[A_{2ij} + B_{2ij} n_f \right] a^4_s(Q) \nonumber\\
& & +\left[D_{2ij} \right] \left(\frac{\pi}{v}\right) a^4_s(Q) \nonumber\\
&=& A_{1ij} a^3_s(Q_1^*) + \left[\tilde{A}_{2ij}\right] a^4_s(Q_1^*) + \nonumber\\
& & D_{2ij} \left(\frac{\pi}{v}\right) a^4_s(Q) ,
\end{eqnarray}
where $(i,j)=(g,q)$ or $(g,\bar{q})$,
\begin{eqnarray}
\ln\frac{Q_1^{*2}}{Q^2} &=& \frac{2B_{2ij}}{A_{1ij}} \nonumber \\
\tilde{A}_{2ij} &=& A_{2ij}+\frac{33}{2}B_{2ij} .\nonumber
\end{eqnarray}

\section{Numerical results and discussions}

The basic input parameters are chosen with the following values: for the top-quark mass, we adopt the present PDG value~\cite{pdg}, which includes the most recent Tevatron Run II result $173.1\pm0.6\pm1.1$ GeV~\cite{cdf}, i.e. $m_t = 172.9\pm1.1$ GeV.

It should be emphasized that the factorization scale $\mu_f$ which enters into the predictions for QCD inclusive reactions is introduced to match nonperturbative and perturbative aspects of the parton distributions in hadrons. The factorization scale should be chosen to match the nonperturbative bound state dynamics with perturbative DGLAP evolution. This could be done by using the nonperturbative models such as AdS/QCD and light-front holography, a recent report can be found in Ref.\cite{ads}. At present, to keep attention on the renormalization scale, we set $\mu_f \equiv m_t$. With such a choice, the dependence on the large logarithms $\ln\mu^2_f/m^2_t$ is eliminated.

To apply PMC scale setting, one has to introduce an initial value for the renormalization scale, i.e. $\mu^{\rm init}_r$. As a reasonable choice for $\mu^{\rm init}_r$, we take $\mu^{\rm init}_{r}=Q$, where $Q$ stands for the typical momentum transfer of the process. For example, $Q$ can be taken as $m_t$, $2m_t$, $\sqrt{s}$, etc. As the default choice, we take $Q=m_t$. In principle, the prediction will not depend on the choice of the initial renormalization scale $\mu^{\rm init}_{r}$ after we have summed the $\{\beta_i\}$-terms up to all orders. There is some residual initial-scale dependence at fixed order because of the unknown-higher-order $\{\beta_i\}$-terms; however we will show that such residual scale dependence can be greatly suppressed by using the PMC.

As is conventional, we have set $\mu_f\equiv\mu_{r}=Q$ in Sec.II. When $\mu_f$ is not equal to $Q$, the formulas listed in the above section should be used with care, i.e. all the $Q$-terms in the scaling functions that exactly correspond to $\mu_f$ should be picked out and fixed during the PMC scale setting. Especially, the terms at NNLO which involve the factor $\ln {Q}^2/\mu^2_{f}$ must be separated into two parts, one is proportional to $\ln\mu^2_f/m_t^2$ which should be kept in its original form, and the other one is proportional to $\ln{Q}^2/m_t^2$ which should be absorbed into the lower-order $\alpha_s$-terms through the standard PMC scale setting. In such a way, we can isolate the correct $n_f$-terms from the NLO and NNLO scaling functions.

For the PDFs, we adopt CTEQ version 2010, i.e. CT10~\cite{cteq}, for our discussion. The effects from the PDF uncertainty and normalization of the QCD coupling will be determined by adopting different PDF sets determined by varying $\alpha_s(m_Z) \in [0.113, 0.230]$. The NNLO $\alpha_s$-running is adopted whose behavior is determined by using the same value of $\alpha_s(m_Z)$ used in the chosen PDF set~\cite{alphas}.

\begin{widetext}
\begin{center}
\begin{table}[ht]
\begin{tabular}{|c||c|c|c|c||c|c|c|c|}
\hline
& \multicolumn{4}{c||}{Conventional scale setting} & \multicolumn{4}{c|}{PMC scale setting} \\
\hline
~~~ ~~~    &~~~LO~~~  &~~~NLO~~~  &~~~NNLO~~~ &~~~ {\it total} ~~~&~~~LO~~~  &~~~NLO~~~  &~~~NNLO~~~ &~~~ {\it total} ~~~\\
\hline
$(q\bar{q})$-channel & 4.989 & 0.975 & 0.489 & 6.453 & 4.841 & 1.756 & -0.063 & 6.489 \\
\hline
$(gg)$-channel    & 0.522 & 0.425 & 0.155 & 1.102 & 0.520 & 0.506  & 0.148  & 1.200 \\
\hline
$(gq)$-channel    & 0.000 &-0.0366 & 0.0050& -0.0316 & 0.000 & -0.0367 & 0.0050  & -0.0315 \\
\hline
$(g\bar{q})$-channel & 0.000 &-0.0367 & 0.0050& -0.0315 & 0.000 & -0.0366 & 0.0050  & -0.0316 \\
\hline
sum     & 5.511  & 1.326 & 0.654 & 7.489 & 5.361 & 2.188  & 0.095  & 7.626 \\
\hline
\end{tabular}
\caption{Total cross-sections (in unit: pb) for the top-quark pair production at the Tevatron with $\sqrt{S}=1.96$ TeV. For the conventional scale setting, we set the renormalization scale $\mu_r\equiv Q$. For the PMC scale setting, we set the initial renormalization scale $\mu^{\rm init}_r=Q$. Here $Q=m_t=172.9$ GeV and the central CT10 as the PDF~\cite{cteq}. }\label{tab1}
\end{table}
\end{center}
\end{widetext}

\begin{widetext}
\begin{center}
\begin{table}[ht]
\begin{tabular}{|c||c|c|c|c||c|c|c|c|}
\hline
& \multicolumn{4}{c||}{Conventional scale setting} & \multicolumn{4}{c|}{PMC scale setting} \\
\hline
~~~ ~~~    &~~~LO~~~  &~~~NLO~~~  &~~~NNLO~~~ &~~~ {\it total} ~~~&~~~LO~~~  &~~~NLO~~~  &~~~NNLO~~~ &~~~ {\it total} ~~~\\
\hline
$(q\bar{q})$-channel & 23.283 & 3.374 & 1.842 & 28.527  & 22.244 & 7.127 & -0.765 & 28.429 \\
\hline
$(gg)$-channel    & 78.692 & 45.918 & 10.637 & 135.113 & 78.399 & 53.570 & 8.539  & 142.548  \\
\hline
$(gq)$-channel    & 0.000 &-0.401 & 1.404 & 1.025    & 0.000  & -0.408 & 1.403 & 1.006 \\
\hline
$(g\bar{q})$-channel & 0.000 &-0.420 & 0.235 & -0.186   & 0.000  & -0.424 & 0.235 & -0.188 \\
\hline
sum   & 101.975 & 48.471 & 14.118  & 164.594 & 100.643 & 59.865 & 9.414 & 171.796\\
\hline
\end{tabular}
\caption{Total cross-sections (in unit: pb) for the top-quark pair production at the LHC with $\sqrt{S}=7$ TeV. For the conventional scale setting, we set the renormalization scale $\mu_r\equiv Q$. For the PMC scale setting, we set the initial renormalization scale $\mu^{\rm init}_r=Q$. Here $Q=m_t=172.9$ GeV and the central CT10 as the PDF~\cite{cteq}. }\label{tab2}
\end{table}
\end{center}
\end{widetext}

\begin{widetext}
\begin{center}
\begin{table}[ht]
\begin{tabular}{|c||c|c|c|c||c|c|c|c|}
\hline
& \multicolumn{4}{c||}{Conventional scale setting} & \multicolumn{4}{c|}{PMC scale setting} \\
\hline
~~~ ~~~    &~~~LO~~~  &~~~NLO~~~  &~~~NNLO~~~ &~~~ {\it total} ~~~&~~~LO~~~  &~~~NLO~~~  &~~~NNLO~~~ &~~~ {\it total} ~~~\\
\hline
$(q\bar{q})$-channel & 73.445 & 9.003 & 5.159 & 87.580 & 69.334 & 21.075 & -2.976 & 86.961 \\
\hline
$(gg)$-channel    & 487.517 & 262.960 & 49.869 & 800.675 & 485.505 & 303.692 & 37.353 & 835.410 \\
\hline
$(gq)$-channel    & 0.000 & 9.299 & 7.685 & 16.996 & 0.000 & 9.369 & 7.687 & 16.967 \\
\hline
$(g\bar{q})$-channel & 0.000 & 0.023 & 1.919 & 1.947 & 0.000 & -0.031 & 1.919 & 17.056 \\
\hline
sum    & 560.962& 281.285& 64.632 & 907.434 & 554.839 & 334.105 & 43.983 & 941.256 \\
\hline
\end{tabular}
\caption{Total cross-sections for the top-quark pair production at the LHC with $\sqrt{S}=14$ TeV. For the conventional scale setting, we set the renormalization scale $\mu_r\equiv Q$. For the PMC scale setting, we set the initial renormalization scale $\mu^{\rm init}_r=Q$. Here $Q=m_t=172.9$ GeV and the central CT10 as the PDF~\cite{cteq}. }\label{tab3}
\end{table}
\end{center}
\end{widetext}

We first present the total cross-sections for the top-quark pair production under the PMC scale setting by fixing all the input parameters to their central values. The results are presented in Tables \ref{tab1}, \ref{tab2} and \ref{tab3}, where for comparison, the total cross-sections for the ``conventional scale setting" method are also presented. Here ``conventional scale setting" means we will directly use Eq.(\ref{subcs}) to do the numerical calculation, which stands for the conventional way of setting the renormalization scale; i.e. $\mu_r\equiv Q$. Three typical hadron collision energies at the Tevatron and LHC are adopted, i.e. $\sqrt{S}=1.96$ TeV, $7$ TeV and $14$ TeV respectively. The results for the four mentioned production channels, i.e. $(q\bar{q})$-, $(gg)$-, $(gq)$- and $(g\bar{q})$-channels, are presented. Note that the result listed in the {\it total}-column is not a simple summation of the corresponding LO, NLO and NNLO results; since they are obtained by using the Sommerfeld re-scattering formula to treat the Coulomb part; i.e., all $(\pi/v)$-terms have been summed up. It is found that such resummed Coulomb term provides an extra $\sim \pm 1\%$ contribution to the total cross-sections.

\begin{itemize}

\item Tables \ref{tab1}, \ref{tab2} and \ref{tab3} show that at the Tevatron the total cross-section is dominated by the $(q\bar{q})$-channel, while at the LHC the dominant channel is the $(gg)$-channel. This is reasonable, since one may observe that the parton luminosity ${\cal L}_{gg} > {\cal L}_{q\bar{q},gq,g\bar{q}}$ in the small $x$-region which is favored at the LHC.

\item Tables \ref{tab1}, \ref{tab2} and \ref{tab3} show that the pQCD convergence is improved after PMC scale setting, especially since the NNLO contribution becomes small. This is due to the fact that we have resummed the universal and gauge invariant higher-order corrections  associated with the $\{\beta_i\}$-terms into the LO and NLO -terms by suitable choice of PMC scales. It is also the reason why after PMC scale setting, the total cross-section $\sigma_{t\bar{t}}$ is increased by $\sim 2\%$ at the Tevatron and $\sim 4\%$ at the LHC. Such a small increment of the total cross-section after PMC scale setting in some sense means that the naive choice of $\mu_r \equiv m_t$ is a viable approximation for estimating the total cross-section. However as will be shown later, by using PMC, the renormalization scale uncertainty is greatly suppressed and essentially eliminated at the NNLO level.

\begin{widetext}
\begin{center}
\begin{figure}
\includegraphics[width=0.40\textwidth]{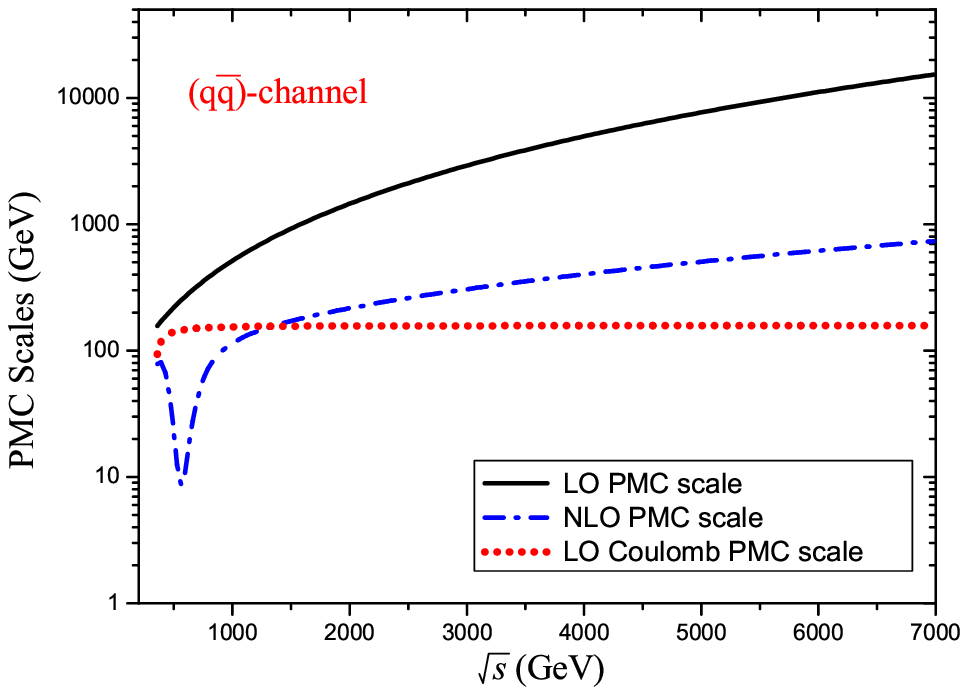}
\hspace{1cm}
\includegraphics[width=0.40\textwidth]{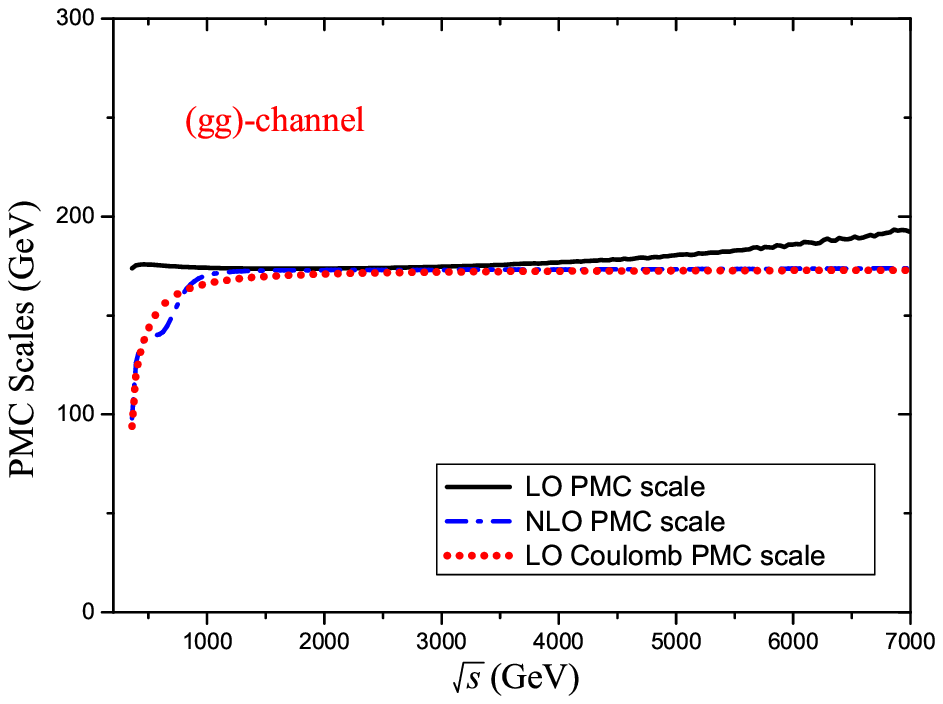}
\caption{PMC scales versus the sub-process collision energy $\sqrt{s}$ for the top-quark pair production up to $\sqrt{s}=7$ TeV, where we have set the initial renormalization scale $\mu^{\rm init}_r=Q$. Here $Q =m_t=172.9$ GeV. }
\label{lhcscale}
\end{figure}
\end{center}
\end{widetext}

\item Since different channels have quite different behaviors, it is necessary to use different PMC scales for each channel. The PMC scales are functions of $\sqrt{s}$, whose behaviors up to $\sqrt{s}=7$ TeV are presented in Fig.(\ref{lhcscale}). Because of the behaviors of the PMC coefficients shown in Fig.(\ref{gqcoe}), the LO PMC scale for the $(q\bar{q})$-channel increases with $\sqrt{s}$ and is much larger than $m_t$ for large $\sqrt{s}$. As a result, its LO cross-sections at the Tevatron and LHC are decreased by $3\%-5\%$ relative to the standard guess of setting $\mu_r\equiv m_t$ under the conventional scale setting method. Because $|B_{1gg}/A_{0gg}|<<1$, the LO PMC scale for the $(gg)$-channel is slightly different from $m_t$ and its LO cross-section remains almost unchanged. It is noted that there is a dip for the NLO scale of the $(q\bar{q})$-channel, which is caused by the fact that $\tilde{B}_{2q\bar{q}}/\tilde{A}_{1q\bar{q}}$ reaches its smallest value when $\sqrt{s} \simeq [\sqrt{2}\exp(5/6)]m_t \sim 563$ GeV. The NLO PMC scale for the $(q\bar{q})$-channel is smaller than $m_t$ by about one order of magnitude in the low $x$-region. As a result, its NLO cross-section will be greatly increased; i.e. it is a factor of two times larger than its value derived without PMC scale setting. As for the $(gg)$-channel, its NLO PMC scale slightly increases with $\sqrt{s}$, but it is smaller than $m_t$ for $\sqrt{s}\lesssim1$ TeV, so that its NLO cross-sections at the Tevatron and LHC are increased by $15\%- 20\%$. These points are shown explicitly in Tables \ref{tab1}, \ref{tab2} and \ref{tab3}.

\begin{center}
\begin{figure}
\includegraphics[width=0.45\textwidth]{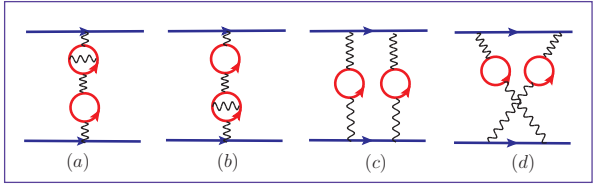}
\caption{Typical $n_f^2$-terms for the electron-muon elastic scattering process at $\alpha^4$-order, where the solid circles stand for the light-lepton loops. Diagrams (a) and (b) are vacuum polarization contributions to the dressed photon propagator which will be absorbed into $\alpha(t)$ as shown by Eq.(\ref{qed}). Diagrams (c) and (d) introduce new type of $\{\beta_i\}$-terms and new PMC scales must be introduced.}
\label{newbeta}
\end{figure}
\end{center}

\item As has been mentioned in the Introduction, there is residual initial renormalization scale dependence because of the unknown-higher-order $\{\beta_i\}$-terms. For the simpler QED process such as the electron-muon elastic scattering through the one-photon exchange only, there is one type of $\{\beta_i\}$-terms, which can be conveniently summed up to all orders and its renormalization scale can be unambiguously set as the virtuality of the exchanged photon as shown by Eq.(\ref{qed}). Thus there is no residual initial scale dependence for the one-photon exchange process. When two or more photon exchange diagrams are involved, more than one types of $\{\beta_i\}$-terms will emerge; i.e. Fig.(\ref{newbeta}c,\ref{newbeta}d) shows the diagrams with two-photon exchange, and there are two types of $\{\beta_i\}$-terms which must be absorbed into two different PMC scales. Because of the unknown higher-order corrections for these two types of $\{\beta_i\}$-terms, there is still residual initial scale dependence.

\begin{widetext}
\begin{center}
\begin{figure}[ht]
\includegraphics[width=0.45\textwidth]{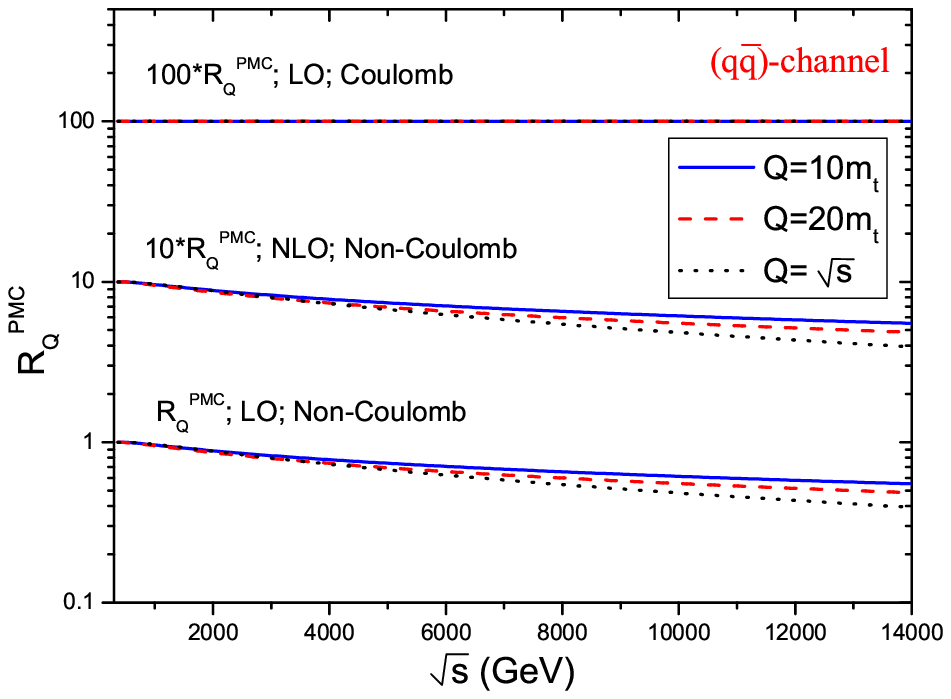}
\includegraphics[width=0.45\textwidth]{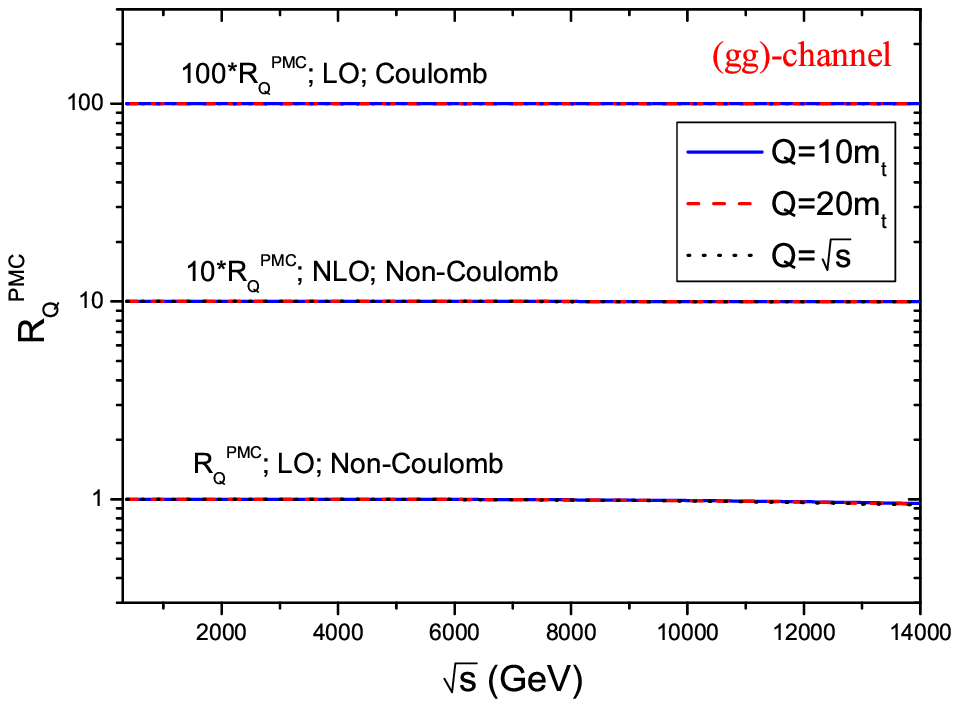}
\caption{The ratio $R_Q^{\rm PMC}= {{\mu^{\rm PMC}_{r}|_{\mu^{\rm init}_{r}=Q}}\over{\mu^{\rm PMC}_{r}|_{\mu^{\rm init}_r = m_t} }} $ versus the sub-process collision energy $\sqrt{s}$ up to $14$ TeV, where $Q=10\,m_t$, $20\,m_t$ and $\sqrt{s}$ respectively. Here $m_t=172.9$ GeV. These results show that the renormalization scales for $t \bar t$ production determined by PMC scale setting at finite order is insensitive to the choice of very disparate initial scales. }
\label{pmcscale}
\end{figure}
\end{center}
\end{widetext}

The PMC scales should be a perturbative series of $\alpha_s$ so as to absorb all $\{\beta_i\}$-dependent terms properly. Therefore, the residual initial renormalization scale uncertainty can be greatly suppressed due to the fact that those higher-order $\{\beta_i\}$-terms can be absorbed into the PMC scales' higher-order terms. We define a ratio $R^{\rm PMC}_{Q}$ to show how the change of initial renormalization scale affects the PMC scales; i.e.
    \begin{equation}
      R_Q^{\rm PMC}=\frac{\mu^{\rm PMC}_r|_{\mu^{\rm init}_{r}=Q}}{\mu^{\rm
      PMC}_r|_{\mu^{\rm init}_{r}=m_t}} ,
    \end{equation}
    where $\mu^{\rm PMC}_r|_{\mu^{\rm init}_{r}=Q}$ stands for the PMC scales determined under the condition of $\mu^{\rm init}_{r}=Q$, which is $Q^*_1$ (LO scale for the non-Coulomb part), $Q^{**}_1$ (NLO scale for the non-Coulomb part) or $Q^{*}_2$ (LO scale for the Coulomb part) respectively. In Fig.(\ref{pmcscale}), we present the value of $R^{\rm PMC}_{Q}$ versus $\sqrt{s}$ up to $14$ TeV. In order to amplify the differences, we take three disparate scales to draw the curves, i.e. $Q=10\,m_t$, $20\,m_t$ and $\sqrt{s}$ respectively. In Fig.(\ref{pmcscale}), the ratios for the dominant $q\bar{q}$- and $gg$- channels are presented, and the ratios for the LO/NLO non-Coulomb PMC scales and LO Coulomb PMC scale are presented in a separate way.

As shown in Fig.(\ref{pmcscale}), the LO PMC scale $Q^{*}_2$ for the Coulomb-term in both channels are unchanged under different choice of Q. Among these choices, $Q=\sqrt{s}$ usually gives the largest deviation from the case of $Q=m_t$. The residual initial scale dependence for the $(gg)$-channel is small, $R_Q^{\rm PMC}\sim 1$, only for the LO non-Coulomb PMC scale $Q^*_1$, it has sizable effect. As an example, for the case of $Q=\sqrt{s}$, its $Q^{*}_1$ deviates from that of $Q=m_t$ by $\sim 1\%$ at $\sqrt{s}=7$ TeV, and it is raised only up to $\sim 7\%$ at $\sqrt{s}=14$ TeV. In the case of the $(q\bar{q})$-channel, the residual scale dependence of the LO/NLO PMC scale for the non-Coulomb part is somewhat larger; i.e. the deviation is about $12\%$ for the case of $Q=\sqrt{s}$ at $\sqrt{s}=2$ TeV, and the deviation reaches up to $\sim 60\%$ at $\sqrt{s}=14$ TeV. (We expect that this dependence on $Q$ will be greatly reduced at NNNLO.) However in such high collision region ($\sqrt{s}>2$ TeV), the total cross-sections are highly suppressed by the parton luminosities and their values are almost unchanged by using very different initial scales.

\begin{widetext}
\begin{center}
\begin{figure}[ht]
\includegraphics[width=0.45\textwidth]{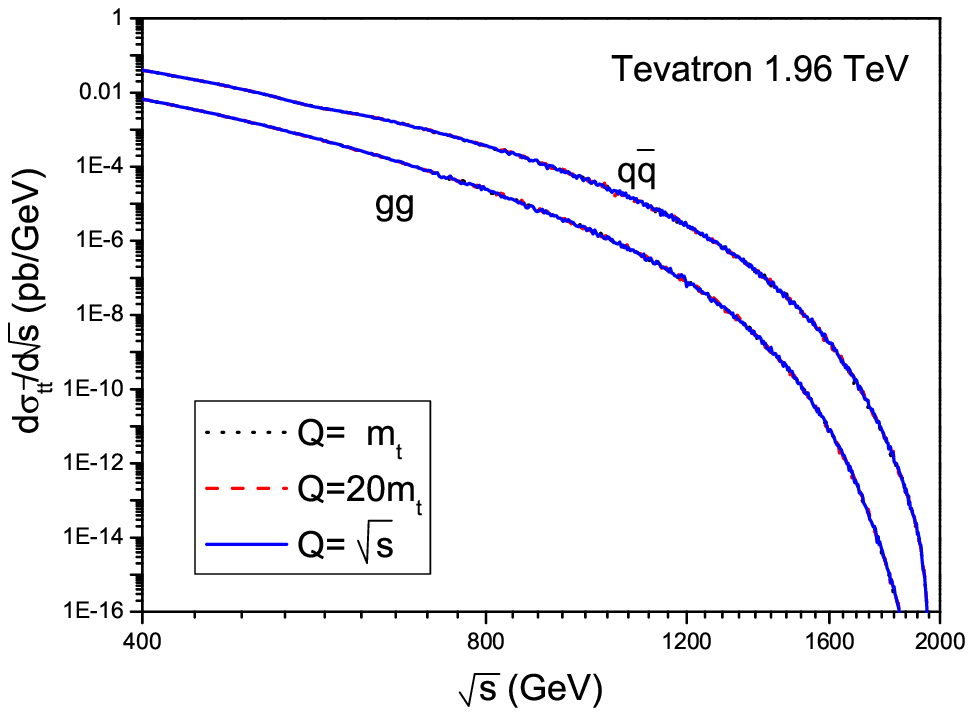}
\includegraphics[width=0.45\textwidth]{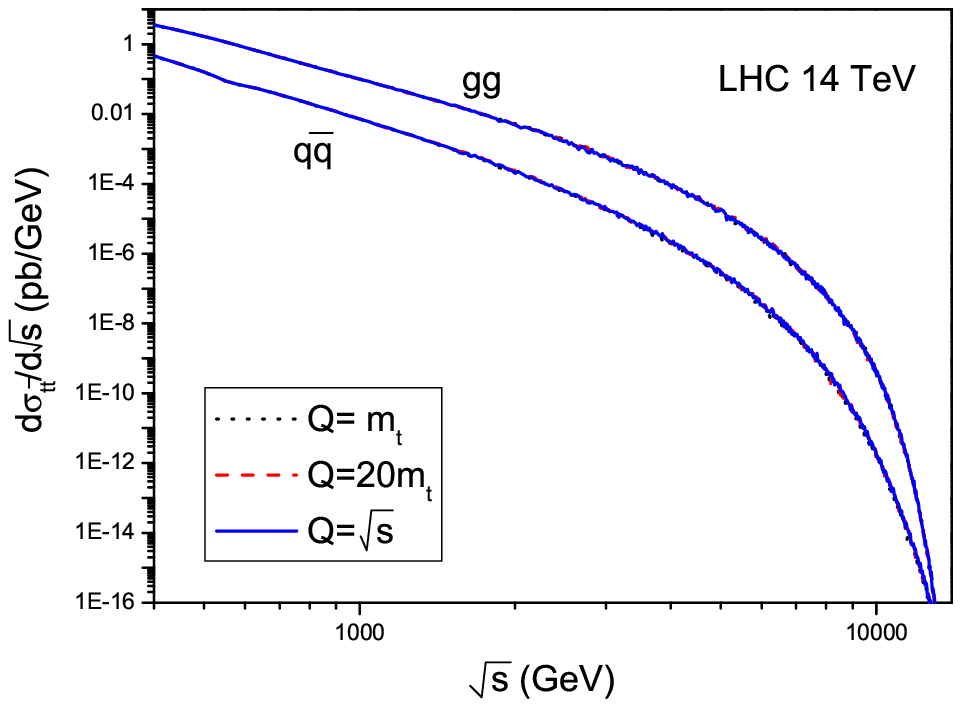}
\caption{A comparison of the hadronic ``effective" differential cross-sections $d\sigma_{t\bar{t}}/d\sqrt{s}\propto\left(2\sqrt{s}{\cal L}_{ij}\hat\sigma_{ij}\right)$ derived from Eq.(5) by taking the initial renormalization scale $Q=m_t$, $20\,m_t$ and $\sqrt{s}$ respectively. The symbol $(ij)$ stands for $(q\bar{q})$ or $(gg)$ respectively. Here $m_t=172.9$ GeV. For each channels, the differential cross-sections with different $Q$ almost coincide with each other. These results thus show that the total cross-sections for $t \bar{t}$ production determined by PMC scale setting at finite order is insensitive to the choice of a wide range of initial renormalization scales. }
\label{diff1}
\end{figure}
\end{center}
\end{widetext}

\begin{widetext}
\begin{center}
\begin{figure}[ht]
\includegraphics[width=0.45\textwidth]{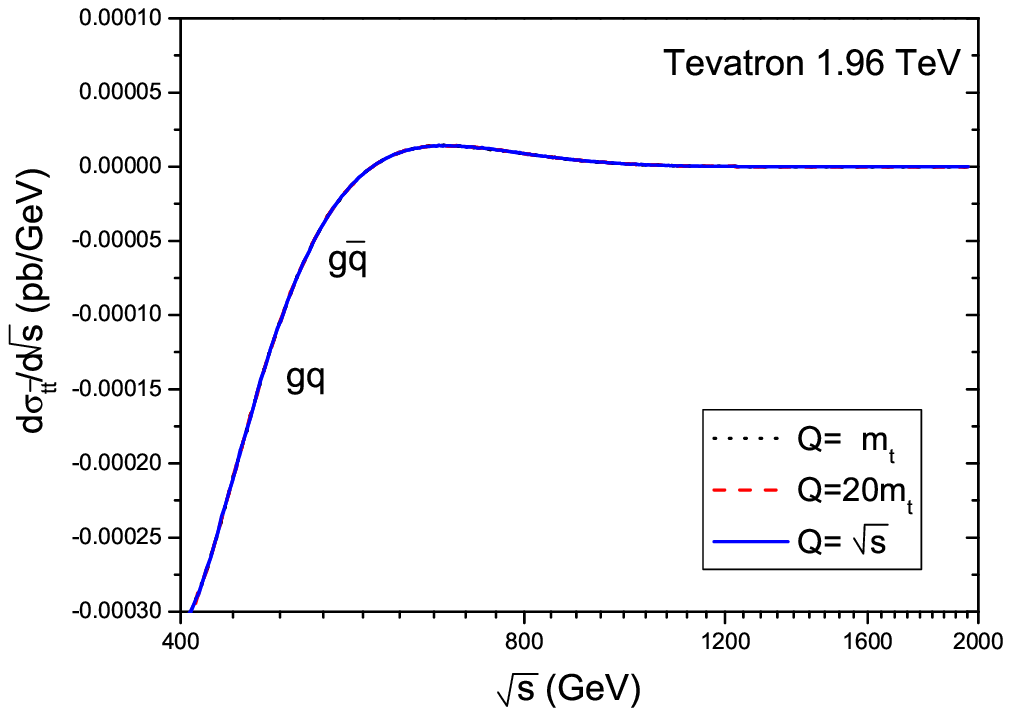}
\includegraphics[width=0.45\textwidth]{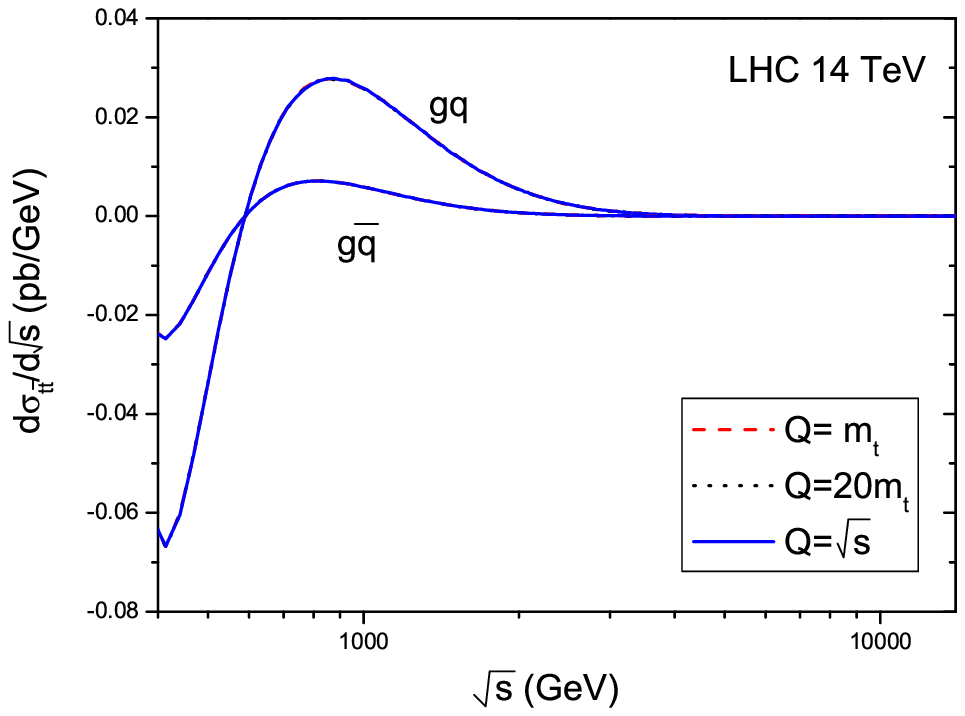}
\caption{A comparison of the hadronic ``effective" differential cross-sections $d\sigma_{t\bar{t}}/d\sqrt{s}\propto\left(2\sqrt{s}{\cal L}_{ij}\hat\sigma_{ij}\right)$ derived from Eq.(5) by taking the initial renormalization scale $Q=m_t$, $20\,m_t$ and $\sqrt{s}$ respectively. The symbol $(ij)$ equals to $(gq)$ or $(g\bar{q})$ respectively. Here $m_t=172.9$ GeV. For each channels, the differential cross-sections with different $Q$ almost coincide with each other. At the Tevatron, the $(gq)$- and $(g\bar{q})$-channels give the same contributions and are negative. These results thus show that the total cross-sections for $t \bar{t}$ production determined by PMC scale setting at finite order is insensitive to the choice of a wide range of initial renormalization scales. }
\label{diff2}
\end{figure}
\end{center}
\end{widetext}

To show this point more clearly, we draw the ``effective" differential cross-sections $d\sigma_{t\bar{t}}/d\sqrt{s}$ versus the subprocess collision energy $\sqrt{s}$ in Fig.(\ref{diff1}) and Fig.(\ref{diff2}), where $d\sigma_{t\bar{t}}/d\sqrt{s}\propto\left(2\sqrt{s}{\cal L}_{ij}\hat\sigma_{ij}\right)$ for a specific $(ij)$-channel which can be derived from Eq.(\ref{basic}). At the Tevatron, the $(gq)$- and $(g\bar{q})$- channels give the same contributions and are negative, so we have put their absolute values in the figure for easy comparison. The differential cross-sections almost coincide with each other. These results thus show that the total cross-sections $\sigma_{t\bar{t}}$ determined by PMC scale setting are insensitive to the difference choices of the initial renormalization scale. More over, Figs.(\ref{diff1},\ref{diff2}) explicitly shows the relative importance of different production channels, i.e. at the Tevatron the total cross-section is dominated by the $(q\bar{q})$-channel, while at LHC the dominant one is the $(gg)$-channel. This agrees with the previous observation from Tables \ref{tab1}, \ref{tab2} and \ref{tab3}.

\begin{widetext}
\begin{center}
\begin{table}[ht]
\begin{tabular}{|c||c|c|c|c|c||c|c|c|}
\hline
& \multicolumn{5}{c||}{PMC scale setting} & \multicolumn{3}{c|}{Conventional scale setting} \\
\hline
~~~~    &~$Q=m_t/4$~ &~$Q=m_t/2$~ &~$Q=m_t$~  &~$Q=2m_t$~ &~$Q=4m_t$~&~$\mu_r\equiv m_t/2$~ &~$\mu_r\equiv m_t$~  &~$\mu_r\equiv 2m_t$~  \\
\hline
Tevatron (1.96 TeV)  & 7.620(5) & 7.622(5) & 7.626(3) & 7.622(6) & 7.623(6) & 7.742(5) & 7.489(3)  & 7.199(5)  \\
\hline
LHC (7 TeV) & 171.6(1) & 171.7(1) & 171.8(1) & 171.7(1) & 171.7(1) & 168.8(1) & 164.6(1)  & 157.5(1) \\
\hline
LHC (14 TeV)& 941.8(8) & 941.9(8) & 941.3(5)  & 941.4(8) & 941.4(8) & 923.8(7) & 907.4(4) & 870.9(6) \\
\hline
\end{tabular}
\caption{Dependence on the initial renormalization scale of the total $t \bar t$ production cross-sections (in unit: pb) at the Tevatron and LHC. Here $m_t=172.9$ GeV and the central CT10 as the PDF~\cite{cteq}. The number in the parenthesis shows the Monte Carlo uncertainty in the last digit. } \label{scaleun}
\end{table}
\end{center}
\end{widetext}

\begin{widetext}
\begin{center}
\begin{table}[ht]
\begin{tabular}{|c||c|c|c|}
\hline
~~~~    &~~$\mu^{\rm PMC}_r\to\mu^{\rm PMC}_r/2$~~  &~~$\mu^{\rm PMC}_r\to\mu^{\rm PMC}_r$~~  &~~$\mu^{\rm PMC}_r\to 2\mu^{\rm PMC}_r$~~  \\
\hline
Tevatron (1.96 TeV)  & 9.683 & 7.626 & 6.163 \\
\hline
LHC (7 TeV) & 223.0 & 171.8 & 136.4 \\
\hline
LHC (14 TeV)& 1221.0 & 941.3 & 748.9 \\
\hline
\end{tabular}
\caption{A comparison of the $t\bar{t}$ production cross-sections (in unit: pb) under ``wrong PMC scales", i.e. all PMC scales $\mu^{\rm PMC}_r$ are changed by $2$ or $1/2$. Here $m_t=172.9$ GeV and the central CT10 as the PDF~\cite{cteq}. Such ``PMC scale uncertainty" shows explicitly how the breaking of conformal symmetry affects the final results.}
\label{pmcscaleun}
\end{table}
\end{center}
\end{widetext}

    Total cross-sections with several typical initial renormalization scale $\mu^{\rm init}_r=Q$ are presented in Table \ref{scaleun}. For the present considered NNLO level, it is found that the residual scale uncertainty to the total cross-section is less than $10^{-3}$ by setting $Q=4m_t$ or $Q =m_t/4$. In fact, even by setting $Q=20\,m_t$ and $\sqrt{s}$, such residual scale uncertainty is still less than $10^{-3}$. As a comparison, we also present the results for the conventional scale setting in Table \ref{scaleun}; by varying the renormalization scale within the region of $[m_t/2,2m_t]$, we obtain a large renormalization scale-uncertainty $\left({}^{+3\%}_{-4\%}\right)$ at the Tevatron and LHC, which agrees with the previous results derived in the literature, c.f. Refs.~\cite{moch2,moch3}. This shows that the renormalization scale uncertainty is greatly suppressed and essentially eliminated using PMC at the NNLO level.

\item Even though the present NNLO result is an approximation, it contains the necessary $\{\beta_i\}$-terms for determining the lower-order PMC scales. After doing the PMC scale setting, the LO- and NLO- terms are conformally invariant and do not depend on the renormalization schemes. However, because the NNNLO $\{\beta_i\}$-contribution is not available at the present, we do not have enough information to set the NNLO PMC scale, so we have set it as $Q^{**}_1$. The NNLO-term is then the only term which contains a non-conformal contribution. However, after PMC scale setting, such remaining non-conformal contribution is negligible; e.g., at the Tevatron it is found that the NNLO contribution itself is reduced to be $\sim 1\%$.

\item After PMC scale setting, the effective PMC scales are unique at each order. Slight change of PMC scales may lead to large effects due to the explicit breaking of the conformal invariance. This, inversely, can be adopted as a check of whether the renormalization scales have been set correctly or not. To see this point more clearly, we make a discussion on the ``PMC scale uncertainty" by varying the PMC scale within the region of $[\mu^{\rm PMC}_r/2,2\mu^{\rm PMC}_r]$, where $\mu^{\rm PMC}_r$ stands for the canonical renormalization scale after PMC scale setting. The results are presented in Table \ref{pmcscaleun}. Large ``PMC scale uncertainty" shows explicitly how the breaking of conformal symmetry affects the final results.

\item We can analyze the combined PDF and $\alpha_s$ uncertainty by using different CTEQ PDF sets, i.e. CT10~\cite{cteq}, which are global fits of experimental data with varying $\alpha_s(m_Z) \in [0.113, 0.230]$. As for the total cross-section after PMC scale setting, we obtain
  \begin{eqnarray}
   \sigma_{\rm Tevatron,\;1.96\,TeV} &=& 7.626^{+0.705}_{-0.610}  \;{\rm pb}\\
   \sigma_{\rm LHC,\;7\,TeV} &=& 171.8^{+19.5}_{-16.2} \;{\rm pb}\\
   \sigma_{\rm LHC,\;14\,TeV} &=& 941.3^{+83.3}_{-77.1} \;{\rm pb}
  \end{eqnarray}
  where the errors are caused by the PDF+$\alpha_s$ uncertainty. Here a larger PDF+$\alpha_s$ error than that of Refs.~\cite{beneke2,moch3} is due to the choice of PDFs with a wider range of $\alpha_s(m_Z)$. If taking the present world average $\alpha_s(m_Z)\simeq 0.118\pm0.001$~\cite{pdg}, we will obtain a much smaller PDF+$\alpha_s$ error; i.e.
\begin{eqnarray}
   \sigma_{\rm Tevatron,\;1.96\,TeV} &=& 7.626^{+0.143}_{-0.130}  \;{\rm pb}\\
   \sigma_{\rm LHC,\;7\,TeV} &=& 171.8^{+3.8}_{-3.5} \;{\rm pb}\\
   \sigma_{\rm LHC,\;14\,TeV} &=& 941.3^{+14.6}_{-15.6} \;{\rm pb}
  \end{eqnarray}

\begin{widetext}
\begin{center}
\begin{figure}
\includegraphics[width=0.40\textwidth]{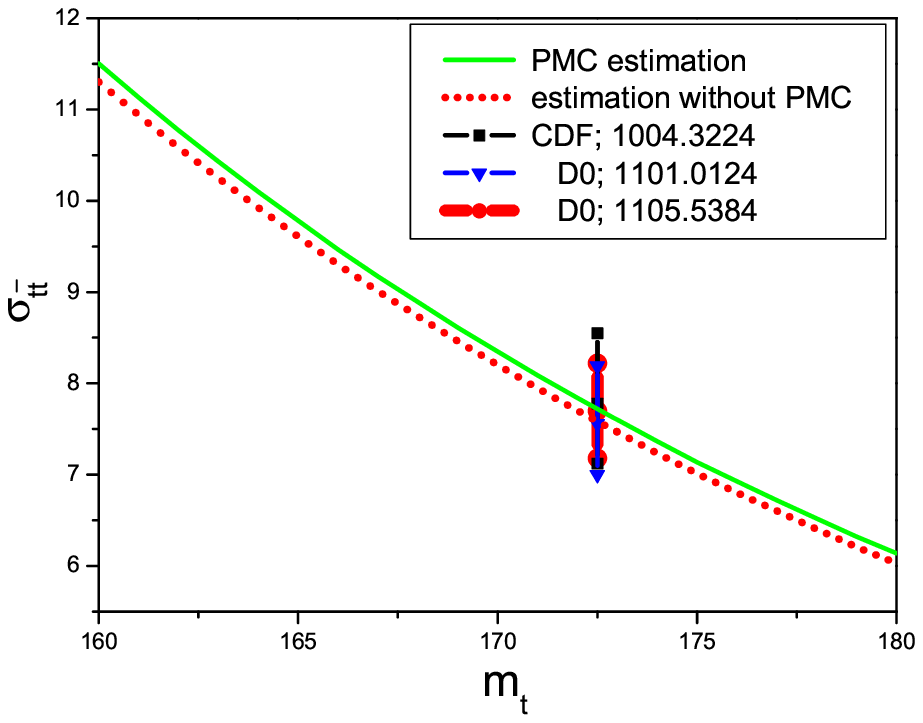}
\includegraphics[width=0.40\textwidth]{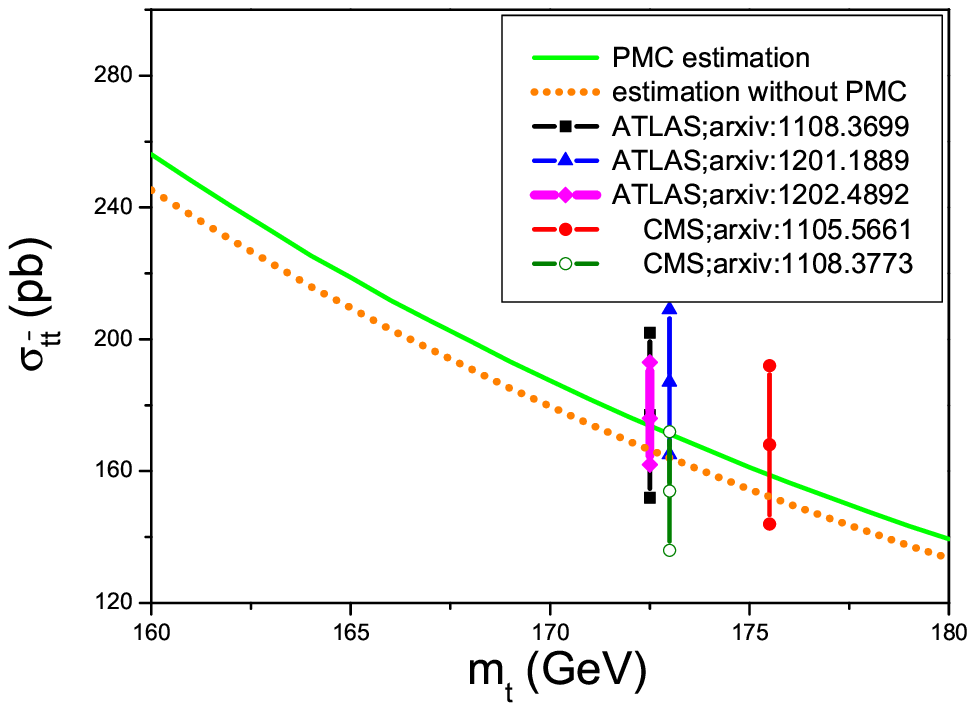}
\caption{Total cross-section $\sigma_{t\bar{t}}$ for the top-pair production, with or without PMC scale setting, versus top-quark mass. The experimental data are adopted from Refs.~\cite{cdf,d0,atlas,cms}. }
\label{mass}
\end{figure}
\end{center}
\end{widetext}

\item The total cross-section $\sigma_{t\bar{t}}$ is sensitive to the top-quark mass, and it is found that the total cross-sections decrease with the increment of top-quark mass. After the PMC scale setting, by varying $m_t=172.9\pm 1.1$ GeV~\cite{pdg}, we predict
  \begin{eqnarray}
   \sigma_{\rm Tevatron,\;1.96\,TeV} &=& 7.626^{+0.265}_{-0.257}  \;{\rm pb}\\
   \sigma_{\rm LHC,\;7\,TeV} &=& 171.8^{+5.8}_{-5.6} \;{\rm pb}\\
   \sigma_{\rm LHC,\;14\,TeV} &=& 941.3^{+28.4}_{-26.5} \;{\rm pb}
  \end{eqnarray}
  where the errors are caused by the top-quark mass uncertainty. In Fig.(\ref{mass}) we present the total cross-section $\sigma_{t\bar{t}}$ as a function of $m_t$, where $\sigma_{t\bar{t}}$ with or without PMC scale setting are shown explicitly. After PMC scale setting, the value of $\sigma_{t\bar{t}}$ is more closer to the central value of the experimental data~\cite{cdf,d0,atlas,cms}, which shows a better agreement with the experimental data.

\end{itemize}

\section{Summary}

The renormalization scales can be set unambiguously by PMC scale setting, which allows us to set the renormalization scale at any required order. The correct scale-displacement between the arguments of different renormalization schemes is automatically set, and the number of active flavors $n_f$ in the $\{\beta_i\}$-function is correctly determined in any scheme. In the present paper, we have presented an explicit example of using PMC scale setting to eliminate the renormalization scale ambiguity and the scheme dependence.

Through PMC scale setting, we have set the LO/NLO PMC scales to the top-quark pair production with the help of the analytic NNLO expressions provided in the literature. After PMC scale setting, all the higher-order $\{\beta_i\}$-terms have been absorbed into the LO- and NLO- $\alpha_s$ running, and we have

\begin{itemize}
\item A larger value for the total $t\bar{t}$ cross-section $\sigma_{t\bar{t}}$ is obtained, which agrees with the present Tevatron and LHC experimental data well. More explicitly, by setting $m_t=172.9\pm 1.1$ GeV, we predict $\sigma_{\rm Tevatron,\;1.96\,TeV} = 7.626^{+0.265}_{-0.257}$ pb, $\sigma_{\rm LHC,\;7\,TeV} = 171.8^{+5.8}_{-5.6}$ pb and $\sigma_{\rm LHC,\;14\,TeV} = 941.3^{+28.4}_{-26.5}$ pb. This is achieved because we have resummed the universal and gauge invariant higher-order corrections which are associated with the running of the coupling into the LO- and NLO- terms by using suitable PMC scales.

\item A more convergent pQCD series expansion is obtained. In contrast to the usual convention of setting $\mu_r\equiv m_t$, the effective LO and NLO PMC scales vary with the subprocess collision energy $\sqrt{s}$. Different channels are distinct and non-interfering, so their PMC scales have been set separately. The procedure is gauge invariant. For the $(q\bar{q})$-channel, its NLO PMC scale is much smaller than $m_t$ in the dominant small $x$-region, and its NLO contribution is largely increased by about a factor of two. In the case of the $(gg)$-channel, its NLO PMC scale slightly increases with increasing $\sqrt{s}$, but it is smaller than $m_t$ for $\sqrt{s}\lesssim1$ TeV, and its NLO contribution is increased by $\sim 20\%$.

\item After PMC scale setting, the resulting LO- and NLO- terms are conformally invariant and do not depend on the choice of renormalization scheme. Since the non-conformal contributions in the NNLO-terms have been suppressed, especially at the Tevatron the NNLO contribution itself is reduced to be $\sim 1\%$. However, slight change of PMC scales will lead to large effects due to the explicit breaking of the conformal invariance. This, inversely, can be adopted as a check of whether the renormalization scales have been set correctly or not. The PDF+$\alpha_s$ uncertainty and the top-quark mass uncertainty have been discussed in detail.

\item In principle, the PMC scale and the resulting renormalized amplitude is independent of the choice of the initial renormalization scale $\mu^{\rm init}_r$. There is residual initial renormalization-scale dependence caused by the lack of information on even higher-order $\{\beta_i\}$-terms. Such residual scale-uncertainty will be greatly suppressed when the PMC scales have been set suitably. At the NNLO level, it is found that the total $t\bar{t}$ cross-sections $\sigma_{t\bar{t}}$ remain almost unchanged by varying the initial scale within a large region of $[m_t/4,4m_t]$, which is shown explicitly in Table \ref{scaleun}. Then, the usual renormalization scale uncertainty $\Delta\sigma_{t\bar{t}}/\sigma_{t\bar{t}} \sim\left({}^{+3\%}_{-4\%}\right)$ at the Tevatron and LHC is greatly suppressed or even eliminated by PMC scale setting. More explicitly, at the LHC with $\sqrt{S}=14$ TeV, it is found that $\sigma_{t\bar{t}}=941 - 942$ pb for $\mu^{\rm init}_{r}\in [m_t/4,4m_t]$, whereas the conventional scale setting leads to a variation of $\sim 6\%$, from $871 - 924$ pb, even when the renormalization scale varies within a smaller region of $[m_t/2,2m_t]$.

\item The determination of the factorization scale is a completely separate issue from the renormalization scale setting. We expect that the factorization scale ambiguity may also be reduced by applying the PMC scale setting in the DGLAP evolution equations.

\item The PMC can be applied to a wide-variety of perturbatively-calculable collider and other processes. Since the renormalization scale and scheme ambiguities are removed, this procedure can improve the precision of tests of the Standard Model and enhance sensitivity to new phenomena.

\end{itemize}

\hspace{1cm}

{\bf Acknowledgements}: We thank Leonardo di Giustino, Robert Shrock, Stefan Hoeche and Michelangelo Mangano for helpful conversations. This work was supported in part by the Program for New Century Excellent Talents in University under Grant NO.NCET-10-0882, Natural Science Foundation of China under Grant NO.11075225, and the Department of Energy contract DE-AC02-76SF00515. SLAC-PUB-14888.

\end{document}